# More powerful biomolecular computers

Janusz Błasiak[1], Tadeusz Krasiński[2], Tomasz Popławski[1], Sebastian Sakowski[2*]

**Biomolecular computers, along with quantum computers, may be a future alternative for traditional, silicon-based computers. Main advantages of biomolecular computers are massive parallel processing of data, expanded capacity of storing information and compatibility with living organisms (first attempts of using biomolecular computers in cancer therapy through blocking of improper genetic information are described in Benenson et al.[1]). However, biomolecular computers have several drawbacks including time-consuming procedures of preparing of input, problems in detecting output signals and interference with by-products. Due to these obstacles, there are few laboratory implementations of theoretically designed DNA computers (like the Turing machine[2,3] and pushdown automaton[4]), but there are many implementations of DNA computers for particular problems [5,6,7,8,9]. The first practical laboratory implementation of the general theoretical model of a machine performing DNA-based calculations was a simple two-symbol two-state finite automaton established by the Shapiro team[10,11]. In the present work, we propose a new attitude, extending the capability of DNA-based finite automaton, by employing two or potentially more restriction enzymes instead of one used in other works. This creates an opportunity to implement in laboratories of more complex finite automata and other theoretical models of computers: pushdown automata, Turing machines.**

A finite automaton represents the simplest model of computer enabling solving plain problems (operations of addition and multiplication of integers cannot be performed by such machines). The finite automaton consists of a tape with cells containing an input word created from symbols of a certain finite alphabet and a control unit reading one after one symbols of the input word and changing its state according to transition rules. Each transition rule is of the form $(s_0, a) \rightarrow s_1$ which means: when an automaton is in the state $s_0$ then after reading the symbol *a* it transits to the state $s_1$. The automaton accepts the input word if starting from the initial state, after reading the whole word, its control unit transits to one of distinguished final states. Usually, a finite automaton is represented by a graph. Figure 1(a) presents an example of two-state automaton which accepts symbols with odd number of *a*'s.

[1] Department of Molecular Genetics, University of Łódź, Pomorska 141/143, 90-236 Łódź, Poland
[2] Faculty of Mathematics and Computer Science, University of Łódź, Banacha 22, 90-238 Łódź, Poland
*E-mail: sebastian.sakowski@op.pl

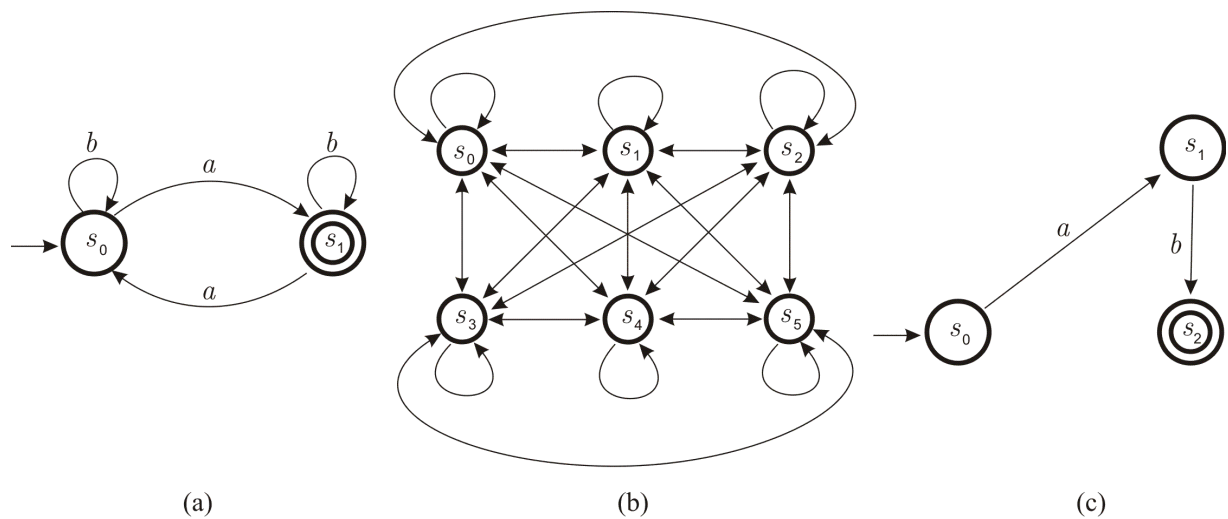

(a)  (b)  (c)

Figure 1. **Finite automata. (a).** Two-state finite automaton accepting strings of symbols with odd number of symbols *a*; $s_0$ – initial state, $s_1$ – final state. **(b).** All possible transitions of the six-state automaton. **(c).** Three-state finite automaton used in the experiment.

The idea of implementation of any two-state, two-symbols finite automaton proposed in[10] was as follows. Two symbols (*a* and *b*), an input word (e.g. *abaa*) and transition rules (e.g. $(s_0, a) \rightarrow s_1$) are encoded by a double-stranded DNA molecules with sticky ends. A 5' single-stranded overhang in the input word represents not only a symbol, but also a state, i.e. a pair <state; symbol> whereas 3' single-stranded fragments in transitions rules contain bases, which are complimentary to bases in the input word. In the automaton computation DNA molecules representing transition rules pair with the input word, the restriction enzyme *Fok*I cleaves the first symbol of the input leaving a sticky end representing the next pair <state, symbol>. Such operations reflects the action of a finite automaton. However, a small number of <state, symbol> codes represented by 4-nt sticky ends left by the action of *Fok*I is a serious disadvantage of such machine. However, it sufficed to encode all possible transitions of the two-state automaton[10]. Attempts of more effective coding or change of restriction enzyme led only to three-state three-symbols automata[12,13].

We propose to employ two restriction enzymes (or potentially more) creating sticky ends of various length. This enables us to extend the capability of the DNA-based finite automaton. Such studies has not been performed so far. In our experiment we used *Acu*I and *Bbv*I restrictases (Figure 2).

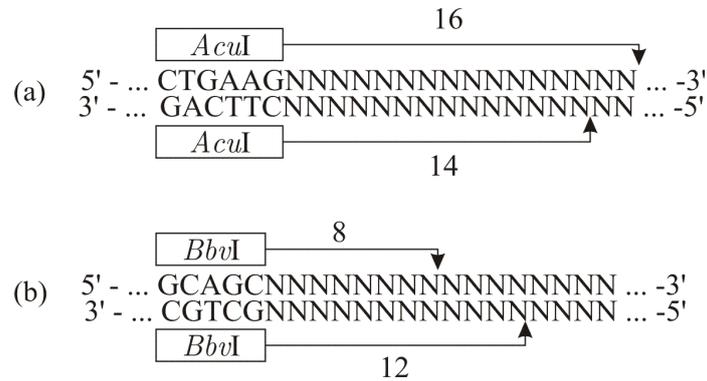

Figure 2. **The action of restriction endonucleases. (a).** *Acu*I, **(b).** *Bbv*I.

The action of these enzymes allowed us to code any six-state two-symbol finite automaton (Figure 1(*b*)). We designed theoretically DNA molecules representing all 72 transitions of the six-state two-symbol finite automaton[14] (see Supplementary Table 1). To test experimentally the idea of using a greater number of restriction enzymes we choose only one automaton, a simple three-state automaton (Figure 1(*c*)). Each of two transitions of the automaton was realized with different enzyme. The DNA codes of the transitions of such automata, an input word *ab* (accepted by this automaton) and an additional molecule indicating the final state are presented in Figure 3.

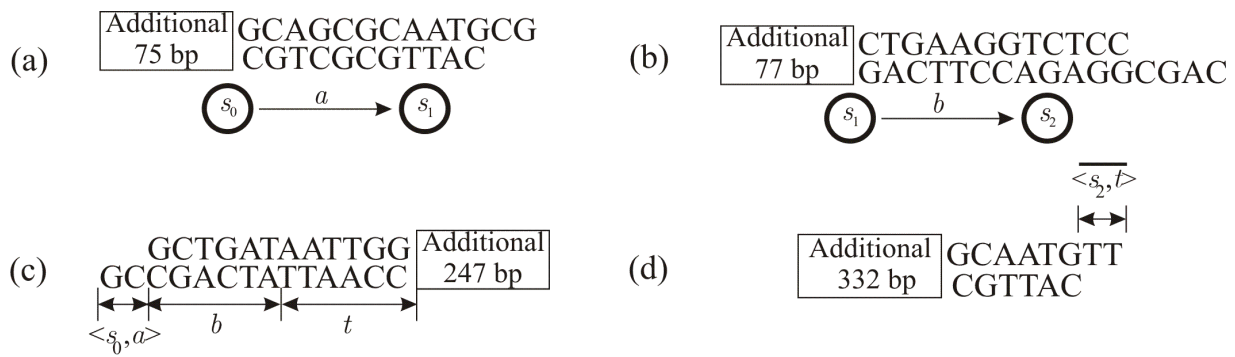

Figure 3. **Components of the experiment. (a).** The transition $(s_0, a) \to s_1$. **(b).** The transition $(s_1, b) \to s_2$. **(c).** The input *ab*. **(d).** The terminal molecule indicating that $s_2$ is the final state.

The action of the whole machine is shown in Figure 4.

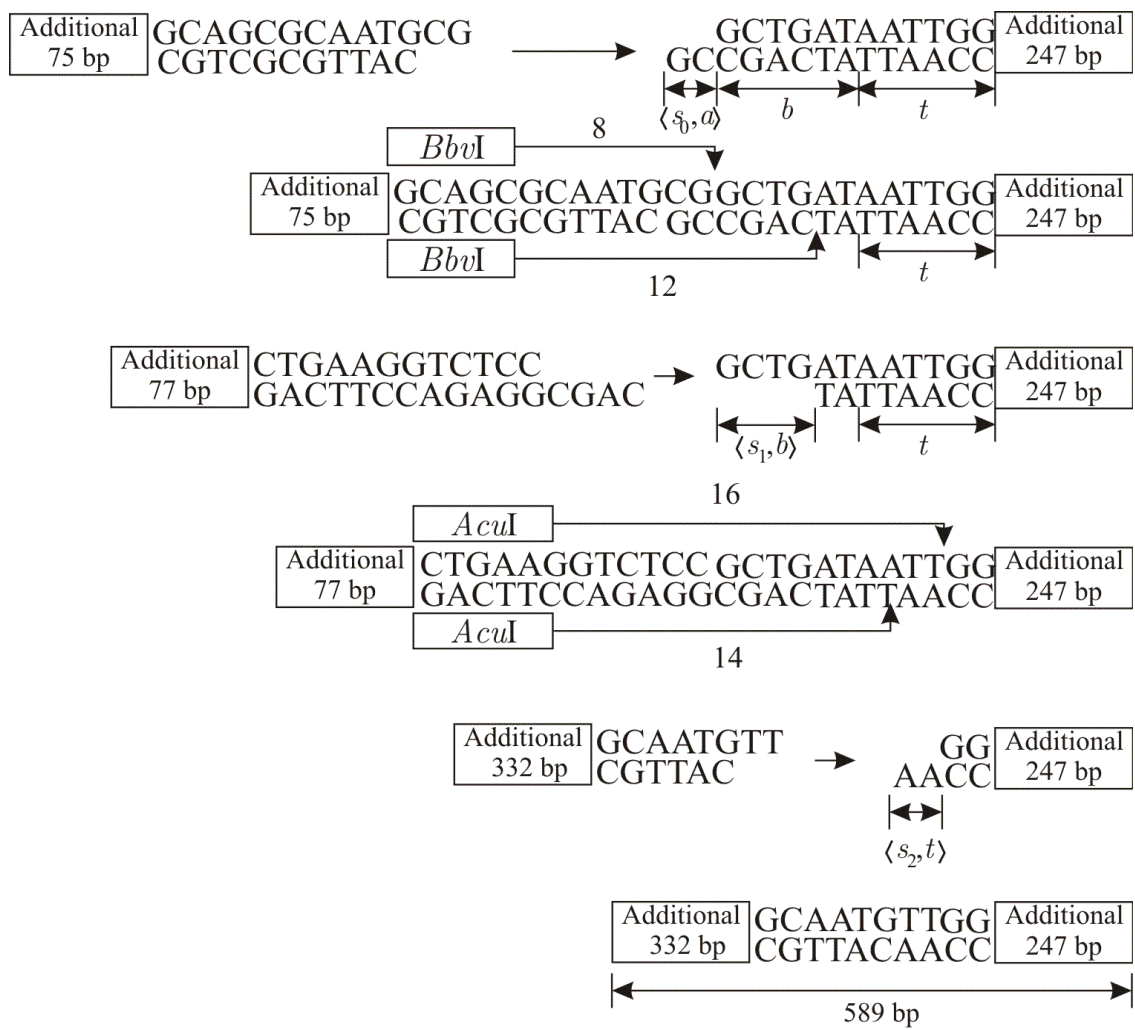

Figure 4. **Computing reactions in the experiment**.

Since the solution contains an additional molecule representing the final state $s_2$, it may hybridize with the rest of the input molecule, producing DNA fragment of 589 bp. Detection of the molecule of such length by gel electrophoresis indicates the acceptance of the input word by the automaton (Figure 5).

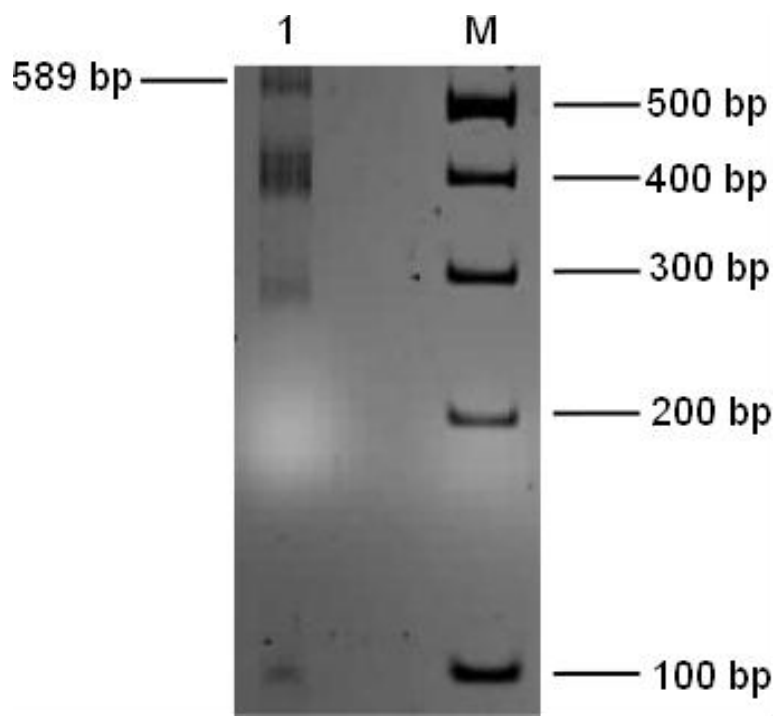

Figure 5. **The result of the experiment**. The composition of two lanes is as follows: 1. the result of computations over the word *ab*. A DNA fragment of 589 length can be seen, indicating the acceptance of the input word *ab* by the automaton. M. the size marker 100 base-pair.

## Materials and methods

**Materials**

Restriction enzymes (REs) *Bbv*I and *Acu*I used as hardware were from New England Biolabs. REs used to prepare DNA molecules representing components of the automaton, *Bsm*AI, *Cla*I, *Bsr*DI, *Not*I and *Acu*I were from New England Biolabs. T4 DNA ligase, T4 polynucleotide kinase (PNL) and pJET 1.2 plasmid were from Fermentas. Synthetic oligonucleotides (liophilized, 200-nmol) were from Metabion. All other chemicals and bacteria media were from Sigma.

**Preparing of automaton's components**

General scheme of preparing automaton's components is different than in the Shapiro team[10]. We propose to built a "DNA library" representing software molecules inserted in plasmids in bacteria (*E. coli*). Once prepared DNA molecules can be used later.

Single-stranded oligonucleotides were labelled AB1, AB2, T661,T662, T671, T672, t1, t2, according to represented components of the automaton (AB1 and complementary AB2 – the input word *ab*, T661 and T662 – the transition T66, T671 and T672 – the transition T67, t1 and t2 – the terminal molecule). They had the following nucleotide sequences: AB1 (5'-TAACTGAAGTCAATCTAAAGTATCGGCTGATAATTGGGAGCAA-3'), AB2 (5'-TTGCTCCCAATTATCAGCCGATACTTTAGATTGACTTCAGTTA-3'). T661 (5'-ACTCAAAGGCGGTAATACGGTTATCCACAGCTGAAGGTCTCCGCTG-3') T662 (5'-CAGCGGAGACCTTCAGCTGTGGATAACCGTATTACCGCCTTTGAGT-3'), T671 (5'-ATCAGGGGATAACGCAGGAAAGAACATGTGCAGCGCAATGCG-3'), T67_2 (5'-CGCATTGCGCTGCACATGTTCTTTCCTGCGTTATCCCCTGAT-3'), t1 (5'- GCGTTTTCCATAGGCTCCGCCCCCCTGACGAGCATCACAAAAATCGACGCTCAAGTCAGAGGTGGCGAAGCAATGTT-3'), t2 (5'-AACATTGCTTCGCCACCTCTGACTTGAGCGTCGATTTTTGTGATGCTCGTCAGGGGGGCGGAGCCTATGGAAAACGC-3').

These oligunucleotides were 5' phosphorylated by T4 PNK. We used 100 pmol of oligonucleotides, which were phosphorylated with 10 u PNK in 20 µl of the PNK buffer and 1 mM ATP, 60 min at 37ºC, precipitated with ethanol and suspended in TE buffer (10 mM Tris HCl, 1 mM EDTA, pH 8.0). Molecules of the automaton's component were prepared by mixing the following pairs of oligonucleotides: AB1 and AB2 (input *ab*); T661 and T662 (transition $(s_0, a) \rightarrow s_1$ ); T671 and T672 (transition $(s_1, b) \rightarrow s_2$ ). The terminal molecule was prepared from t1 and t2. Complimentary oligonucleotides (sense and antisense) were incubated 10 min at 37ºC, and then slowly (1ºC/min) cooled to room temperature. Resulting double-stranded DNA fragments were cloned with the pJET 1.2 plasmid vector and then sequenced in Institute of Biochemistry and Biophysics, Warsaw, Poland. Clones labelled as pPSAB, pPSTER, pPS66, pPS67 were chosen for further experiments. Proper DNA fragments were obtained by the cleavage of pPSAB, pPSTER, pPS66, pPS67 with restriction enzyme, agarose gel electrophoresis and purification with the QIAquick DNA Purification Kit. The DNA fragment representing input *x=ab* was obtained with the *Acu*I enzyme, the transition T66 with *Bsm*AI and *Cla*I enzymes, transition T67 with *Not*I and *Bsr*DI, Term with *Bsr*DI and *Bsm*AI.

**Computation reactions**

This reaction was run for 1.5 h in NEB4 buffer at 37ºC. Reaction tube contained a set of DNA fragments representing input and transitions molecules, 1u *Bbv*I and *Acu*I as well as 10 u T4 DNA ligase. The product of the reaction was purified three times with phenol, chloroform and izoamyl alcohol (25:24:1) and sorted by 8% polyacrylamide gel electrophoresis.

**Conclusions**

Our experiment has indicated a new direction in the extending of DNA-based finite automaton by employing two restriction enzymes acting in one tube. This creates an opportunity to implement more complex finite automata and other theoretical models of computers like pushdown automata or Turing machines by using potentially more restriction enzymes. We also proposed to built DNA library of inputs and transition rules, which can be used as needed. In the future we plan to determine the influence of: the length of input, number of states and non-determinism of the automaton on the functioning of finite automata.

**Acknowledgements**

The authors acknowledge support from the Rector of the University of Łódź (Grant no: 505/376).

# Supplementary information

# More powerful biomolecular computers

Janusz Błasiak[1], Tadeusz Krasiński[2], Tomasz Popławski[1], Sebastian Sakowski[2*]

[1] Department of Molecular Genetics, University of Łódź, Pomorska 141/143, 90-236 Łódź, Poland

[2] Faculty of Mathematics and Computer Science, University of Łódź, Banacha 22, 90-238 Łódź, Poland

**Table of Contents:**

**Table 1**. The codes of all transitions of 6-state 2-symbol finite automaton.

[*]Contact: sebastian.sakowski@op.pl

**Table 1**

**The codes of all transitions of 6-state 2-symbol finite automaton.**

| TRANSITONS | CODES | TRANSITONS | CODES |
|---|---|---|---|
| **T1:** $s_0 \xrightarrow{a} s_0$ | 5'-GCAGCNN    -3'<br>3'-CGTCGNNCAGC-5' | **T19:** $s_3 \xrightarrow{a} s_0$ | 5'-GCAGCNNNNCG-3'<br>3'-CGTCGNNNN  -5' |
| **T2:** $s_0 \xrightarrow{a} s_1$ | 5'-GCAGCNNN   -3'<br>3'-CGTCGNNNCAGC-5' | **T20:** $s_1 \xrightarrow{a} s_3$ | 5'-CTGAAGNNNNN    -3'<br>3'-GACTTCNNNNNTCAG-5' |
| **T3:** $s_0 \xrightarrow{a} s_2$ | 5'-GCAGCNNNN   -3'<br>3'-CGTCGNNNNCAGC-5' | **T21:** $s_1 \xrightarrow{a} s_4$ | 5'-CTGAAGNNNNNN   -3'<br>3'-GACTTCNNNNNNTCAG-5' |
| **T4:** $s_0 \xrightarrow{b} s_0$ | 5'-GCAGCNN   -3'<br>3'-CGTCGNNACTA-5' | **T22:** $s_1 \xrightarrow{a} s_5$ | 5'-CTGAAGNNNNNNN   -3'<br>3'-GACTTCNNNNNNNTCAG-5' |
| **T5:** $s_0 \xrightarrow{b} s_1$ | 5'-GCAGCNNN   -3'<br>3'-CGTCGNNNACTA-5' | **T23:** $s_1 \xrightarrow{b} s_3$ | 5'-CTGAAGNNNNN   -3'<br>3'-GACTTCNNNNNGACT-5' |
| **T6:** $s_0 \xrightarrow{b} s_2$ | 5'-GCAGCNNNN    -3'<br>3'-CGTCGNNNNACTA-5' | **T24:** $s_1 \xrightarrow{b} s_4$ | 5'-CTGAAGNNNNNN   -3'<br>3'-GACTTCNNNNNNGACT-5' |
| **T7:** $s_3 \xrightarrow{a} s_3$ | 5'-CTGAAGNNNNNNNNCG-3'<br>3'-GACTTCNNNNNNNN  -5' | **T25:** $s_1 \xrightarrow{b} s_5$ | 5'-CTGAAGNNNNNNN   -3'<br>3'-GACTTCNNNNNNNGACT-5' |
| **T8:** $s_3 \xrightarrow{a} s_4$ | 5'-CTGAAGNNNNNNNNNCG-3'<br>3'-GACTTCNNNNNNNNN  -5' | **T26:** $s_3 \xrightarrow{a} s_1$ | 5'-GCAGCNNNNNCG-3'<br>3'-CGTCGNNNNN  -5' |
| **T9:** $s_3 \xrightarrow{a} s_5$ | 5'-CTGAAGNNNNNNNNNNCG-3'<br>3'-GACTTCNNNNNNNNNN  -5' | **T27:** $s_4 \xrightarrow{a} s_1$ | 5'-GCAGCNNNNTC-3'<br>3'-CGTCGNNNN  -5' |
| **T10:** $s_3 \xrightarrow{b} s_3$ | 5'-CTGAAGNNNNNNNNAT-3'<br>3'-GACTTCNNNNNNNN  -5' | **T28:** $s_5 \xrightarrow{a} s_1$ | 5'-GCAGCNNNGT-3'<br>3'-CGTCGNNN  -5' |
| **T11:** $s_3 \xrightarrow{b} s_4$ | 5'-CTGAAGNNNNNNNNNAT-3'<br>3'-GACTTCNNNNNNNNN  -5' | **T29:** $s_3 \xrightarrow{b} s_1$ | 5'-GCAGCNNNNNAT-3'<br>3'-CGTCGNNNNN  -5' |
| **T12:** $s_3 \xrightarrow{b} s_5$ | 5'-CTGAAGNNNNNNNNNNAT-3'<br>3'-GACTTCNNNNNNNNNN  -5' | **T30:** $s_4 \xrightarrow{b} s_1$ | 5'-GCAGCNNNNGA-3'<br>3'-CGTCGNNNN  -5' |
| **T13:** $s_0 \xrightarrow{a} s_3$ | 5'-CTGAAGNNNNN   -3'<br>3'-GACTTCNNNNNCAGC-5' | **T31:** $s_5 \xrightarrow{b} s_1$ | 5'-GCAGCNNNTG-3'<br>3'-CGTCGNNN  -5' |
| **T14:** $s_0 \xrightarrow{a} s_4$ | 5'-CTGAAGNNNNNN   -3'<br>3'-GACTTCNNNNNNCAGC-5' | **T32:** $s_2 \xrightarrow{a} s_2$ | 5'-GCAGCNN   -3'<br>3'-CGTCGNNATCA-5' |
| **T15:** $s_0 \xrightarrow{a} s_5$ | 5'-CTGAAGNNNNNNN   -3'<br>3'-GACTTCNNNNNNNCAGC-5' | **T33:** $s_2 \xrightarrow{a} s_1$ | 5'-GCAGCN   -3'<br>3'-CGTCGNATCA-5' |
| **T16:** $s_0 \xrightarrow{b} s_3$ | 5'-CTGAAGNNNNN   -3'<br>3'-GACTTCNNNNNNACTA-5' | **T34:** $s_2 \xrightarrow{a} s_0$ | 5'-GCAGC   -3'<br>3'-CGTCGATCA-5' |
| **T17:** $s_0 \xrightarrow{b} s_4$ | 5'-CTGAAGNNNNNN   -3'<br>3'-GACTTCNNNNNNACTA-5' | **T35:** $s_2 \xrightarrow{b} s_2$ | 5'-GCAGCNN   -3'<br>3'-CGTCGNNCGAC-5' |
| **T18:** $s_0 \xrightarrow{b} s_5$ | 5'-CTGAAGNNNNNNN   -3'<br>3'-GACTTCNNNNNNNACTA-5' | **T36:** $s_2 \xrightarrow{b} s_1$ | 5'-GCAGCN   -3'<br>3'-CGTCGNCGAC-5' |

| TRANSITONS | CODES | TRANSITONS | CODES |
|---|---|---|---|
| **T37:** $s_2 \xrightarrow{b} s_0$ | `5'-GCAGC     -3'`<br>`3'-CGTCGCGAC-5'` | **T55:** $s_4 \xrightarrow{b} s_5$ | `5'-CTGAAGNNNNNNNNGA-3'`<br>`3'-GACTTCNNNNNNNN   -5'` |
| **T38:** $s_5 \xrightarrow{a} s_5$ | `5'-CTGAAGNNNNNNNNGT-3'`<br>`3'-GACTTCNNNNNNNN   -5'` | **T56:** $s_5 \xrightarrow{a} s_4$ | `5'-CTGAAGNNNNNNGT-3'`<br>`3'-GACTTCNNNNNNN  -5'` |
| **T39:** $s_4 \xrightarrow{a} s_0$ | `5'-GCAGCNNNTC-3'`<br>`3'-CGTCGNNN   -5'` | **T57:** $s_5 \xrightarrow{a} s_3$ | `5'-CTGAAGNNNNNGT-3'`<br>`3'-GACTTCNNNNNN  -5'` |
| **T40:** $s_5 \xrightarrow{a} s_0$ | `5'-GCAGCNNGT-3'`<br>`3'-CGTCGNN  -5'` | **T58:** $s_5 \xrightarrow{b} s_5$ | `5'-CTGAAGNNNNNNNNTG-3'`<br>`3'-GACTTCNNNNNNNN   -5'` |
| **T41:** $s_3 \xrightarrow{a} s_0$ | `5'-GCAGCNNNNAT-3'`<br>`3'-CGTCGNNNN   -5'` | **T59:** $s_5 \xrightarrow{b} s_4$ | `5'-CTGAAGNNNNNNNTG-3'`<br>`3'-GACTTCNNNNNNN  -5'` |
| **T42:** $s_4 \xrightarrow{b} s_0$ | `5'-GCAGCNNNGA-3'`<br>`3'-CGTCGNNN   -5'` | **T60:** $s_5 \xrightarrow{b} s_3$ | `5'-CTGAAGNNNNNNTG-3'`<br>`3'-GACTTCNNNNNNN  -5'` |
| **T43:** $s_5 \xrightarrow{b} s_0$ | `5'-GCAGCNNTG-3'`<br>`3'-CGTCGNN  -5'` | **T61:** $s_2 \xrightarrow{a} s_3$ | `5'-CTGAAGNNNN      -3'`<br>`3'-GACTTCNNNNATCA-5'` |
| **T44:** $s_1 \xrightarrow{a} s_1$ | `5'-GCAGCNN     -3'`<br>`3'-CGTCGNNTCAG-5'` | **T62:** $s_2 \xrightarrow{a} s_4$ | `5'-CTGAAGNNNNN     -3'`<br>`3'-GACTTCNNNNNATCA-5'` |
| **T45:** $s_1 \xrightarrow{a} s_0$ | `5'-GCAGCN     -3'`<br>`3'-CGTCGNTCAG-5'` | **T63:** $s_2 \xrightarrow{a} s_5$ | `5'-CTGAAGNNNNNN     -3'`<br>`3'-GACTTCNNNNNNATCA-5'` |
| **T46:** $s_1 \xrightarrow{a} s_2$ | `5'-GCAGCNNN     -3'`<br>`3'-CGTCGNNNTCAG-5'` | **T64:** $s_2 \xrightarrow{b} s_3$ | `5'-CTGAAGNNNN     -3'`<br>`3'-GACTTCNNNNCGAC-5'` |
| **T47:** $s_1 \xrightarrow{b} s_1$ | `5'-GCAGCNN     -3'`<br>`3'-CGTCGNNGACT-5'` | **T65:** $s_2 \xrightarrow{b} s_4$ | `5'-CTGAAGNNNNN     -3'`<br>`3'-GACTTCNNNNNCGAC-5'` |
| **T48:** $s_1 \xrightarrow{b} s_0$ | `5'-GCAGCN     -3'`<br>`3'-CGTCGNGACT-5'` | **T66:** $s_2 \xrightarrow{b} s_5$ | `5'-CTGAAGNNNNNN     -3'`<br>`3'-GACTTCNNNNNNCGAC-5'` |
| **T49:** $s_1 \xrightarrow{b} s_2$ | `5'-GCAGCNNN     -3'`<br>`3'-CGTCGNNNGACT-5'` | **T67:** $s_3 \xrightarrow{a} s_2$ | `5'-GCAGCNNNNNNCG-3'`<br>`3'-CGTCGNNNNNN   -5'` |
| **T50:** $s_4 \xrightarrow{a} s_4$ | `5'-CTGAAGNNNNNNNTC-3'`<br>`3'-GACTTCNNNNNNN  -5'` | **T68:** $s_4 \xrightarrow{a} s_2$ | `5'-GCAGCNNNNNTC-3'`<br>`3'-CGTCGNNNNN   -5'` |
| **T51:** $s_4 \xrightarrow{a} s_3$ | `5'-CTGAAGNNNNNNTC-3'`<br>`3'-GACTTCNNNNNNN  -5'` | **T69:** $s_5 \xrightarrow{a} s_2$ | `5'-GCAGCNNNNGT-3'`<br>`3'-CGTCGNNNN   -5'` |
| **T52:** $s_4 \xrightarrow{a} s_5$ | `5'-CTGAAGNNNNNNNNTC-3'`<br>`3'-GACTTCNNNNNNNN   -5'` | **T70:** $s_3 \xrightarrow{b} s_2$ | `5'-GCAGCNNNNNNAT-3'`<br>`3'-CGTCGNNNNNN   -5'` |
| **T53:** $s_4 \xrightarrow{b} s_4$ | `5'-CTGAAGNNNNNNNGA-3'`<br>`3'-GACTTCNNNNNNNN  -5'` | **T71:** $s_4 \xrightarrow{b} s_2$ | `5'-GCAGCNNNNGA-3'`<br>`3'-CGTCGNNNNN   -5'` |
| **T54:** $s_4 \xrightarrow{b} s_3$ | `5'-CTGAAGNNNNNNGA-3'`<br>`3'-GACTTCNNNNNNN  -5'` | **T72:** $s_5 \xrightarrow{b} s_2$ | `5'-GCAGCNNNNTG-3'`<br>`3'-CGTCGNNNN   -5'` |